# The Far-Infrared Surveyor Mission Study: Paper I, the Genesis

M.Meixner[*1a,b], A. Cooray[c],R. Carter[d], M. DiPirro[d], A. Flores[d], D. Leisawitz[d], L. Armus[e], C. Battersby[f], E. Bergin[g], C.M. Bradford[h], K. Ennico[i], G. J. Melnick[f], S. Milam[d], D. Narayanan[j], K. Pontoppidan[a], A. Pope[k], T. Roellig[i], K. Sandstrom[l], K.Y.L. Su[m], J. Vieira[n], E. Wright[o], J. Zmuidzinas[p], S. Alato[q], S. Carey[e], M. Gerin[r], F. Helmich[s], K. Menten[t], D. Scott[u], I. Sakon[v], R. Vavrek[w]

[a]Space Telescope Science Institute, 3700 San Martin Dr., Baltimore, MD 21218, [b]Dept. of Physics and Astronomy, The Johns Hopkins University, 3400 N. Charles St., Baltimore, MD 21218, [c]Dept. of Physics and Astronomy, University of California, Irvine, [d] NASA Goddard Space Flight Center, 8800 Greenbelt Rd., Greenbelt, MD 20771 USA, [e]NASA Infrared Processing and Analysis Center,[f]Harvard-Smithsonian Center for Astrophysics,[g]Dept. of Astronomy, University of Michigan,[h]NASA Jet Propulsion Laboratory,[i]NASA Ames Research Center,[j]University of Florida, Gainsville,[k]Department of Astronomy, University of Massachusetts, LGRT-B 619E, Amherst, MA 01003, USA,[l]University of California, San Diego,[m]Steward Observatory, University of Arizona,[n]University of Illinois, Urbana-Champaign,[o]University of California, Los Angeles, [p]California Institute of Technology,[q]SNSB, [r]CNES, [s]SRON, [t]DLR, [u]CSA, [v]JAXA, [w]ESA

## ABSTRACT

This paper describes the beginning of the Far-Infrared Surveyor mission study for NASA's Astrophysics Decadal 2020. We describe the scope of the study, and the open process approach of the Science and Technology Definition Team. We are currently developing the science cases and provide some preliminary highlights here. We note key areas for technological innovation and improvements necessary to make a Far-Infrared Surveyor mission a reality.

## 1. INTRODUCTION

The Far-Infrared (Far-IR) Surveyor concept originates most recently in the NASA 2013 Roadmap[1], but has a history within the community. This mission builds upon decades of success in space/airborne infrared astronomy with IRAS[2], KAO[3], ISO[4], Spitzer[5], Herschel[6], SOFIA[7] and soon to be launched, JWST[8]. The Roadmap envisaged a space observatory with a large gain in sensitivity over the Herschel Space Observatory[6], better angular resolution sufficient to overcome spatial confusion in deep cosmic surveys and new spectroscopic capabilities. A white paper from the 2015 Pasadena Far-Infrared Community workshop[9], provides a recent starting point that summarizes some key questions in the field and a description of a potential, single aperture telescope.

The Science and Technology Definition Team (STDT) were selected to represent the US astronomy community's demographics, geography, and expertise areas and include: Margaret Meixner, Asantha Cooray, David Leisawitz, Lee Armus, Cara Battersby, Ted Bergin, Matt Bradford, Kimberly Ennico, Gary Melnick, Stefanie Milam, Desika Narayanan, Klaus Pontoppidan, Alexandra Pope, Tom Roellig, Karin Sandstrom, Kate Su, Joaquin Vieira, Ed Wright, and Jonas Zmuidinas. This group, which is at the nexus of the mission concept study, is co-led by the two community co-chairs, Asantha Cooray and Margaret Meixner, who are responsible for delivery of the reports to NASA headquarters (HQ) and the Decadal Committee. The STDT community co-chairs direct the engineering study that will be performed by NASA Goddard Space Flight Center with study manager, Ruth Carter, study scientist, David Leisawitz, study technologist, Mike DiPirro, and study mission systems engineer, Anel Flores. However, we anticipate assistance from other NASA centers including Ames, JPL and Marshall. The STDT draws upon input from the astronomical community through the Far-IR Science Interest Group (SIG), NASA government labs, (Goddard, JPL, Ames), Academia, and

---

Industry.  The NASA HQ Program Scientists, Kartik Sheth and Dominic Benford,  NASA Cosmic Origins Program Scientists, Susan Neff and Deborah Padgett,  and IPAC support staff scientist, Sean Carey, observe and support the study as ex-officio, non-voting representatives. Several ex-officio non-voting international organizational representatives are observing and contributing to the process including: Susanne Alato (SNSB), Maryvonne Gerin (CNES), Frank Helmich (SRON), Karl Menten (DLR), Douglas Scott (CAS), Itsuki Sakon (JAXA), Roland Vavrek (ESA). Finally, the STDT chairs have established a Senior Advisory board with past experience with the Decadal process, to be a technical sounding board and a review board for drafts of the report. This senior advisory board includes Marcia Rieke, Jean Turner, Sarah Lipscy, Harry Ferguson, Charles Lawrence, Harvey Moseley, George Helou, John Carlstrom, Jon Arenberg, Meg Urry, George Rieke and John Mather.

The STDT relies on a number of working groups to prioritize the science identification and science drivers of the mission architecture. The work related to mission concept development is coordinated through a separate working group whose primary expertise on issues involving technology. While members of the STDT lead each of the working groups they are primarily composed of international community members. The five science working groups (with leads identified) are: Reionization and Cosmology (Matt Bradford, Joaquin Vieira), Evolution of Galaxies and Blackholes (Alexandra Pope and Lee Armus),  Milky Way, ISM and Local Volume of Galaxies (Cara Battersby, Karin Sandstrom), Protoplanetary disks and Exoplanets (Klaus Pontoppidan, Kate Su),  and Solar System (Stefanie Milam). The Mission Concept Development Group is coordinated by Tom Roellig. The mission development relies on the identification of primary science drivers and establishing the technical requirements (e.g., spatial and spectral resolution, continuum or spectral sensitivity, survey area or number of targets, wavelength coverage, among others). Key science themes identified by the science working groups are outlined in Section 2.

The main deliverable is a report for the Decadal 2020 committee that provides a concept maturity Level 4 design for the Far-IR Surveyor from the engineering study and the science case which justifies the cost of the mission. There will be interim reports and deliverables along the way. The process is intended to be open and transparent to the community which is international in nature. Biweekly telecons are open to everyone and talks from the face-to-face meetings are posted on our public websites: asd.gsfc.nasa.gov/firs.

## 2.  SCIENCE CASE

The science pursued by such a Far-Infrared Surveyor must be posed in the 2030 context. This scientific landscape will have been shaped by 15+ years of Atacama Large Millimeter Array (ALMA) observations, the discoveries by the James Webb Space Telescope (JWST), the Stratospheric Observatory for Infrared Astronomy (SOFIA), and the Wide-Field Infrared Survey Telescope (WFIRST) and the results from newly operational 25-35 m ground based telescope facilities. The Far-IR Surveyor is envisioned to be a general purpose space observatory with a broad science case that we highlight in five areas below.

### 2.1 Cosmic Dawn and Reionization

Rest-frame UV and optical lines allow studies on the ISM and gas-phase metallicities. With *Hubble*, such studies have been attempted at $z \sim 1$ to 2 and will soon be extended to $z \sim 6$ with JWST. In the post-JWST era a far-infrared space telescope with at least a four-order of magnitude sensitivity improvement over Herschel will enable studies on the gas properties, AGN activity and star- formation within galaxies at redshifts $z = 6$ to 12. This redshift range covering the epoch of reionization is especially important for our understanding of the cosmic origins, formation of first stars, galaxies and blackholes, and the onset of large-scale structure we see today. In particular, 20 to 600 μm spectroscopic observations can: (1) disentangle the complex conditions in the ISM of primordial galaxies by measuring the gas densities and excitation, and the prevalence of shock heating; (2) use spectral line diagnostics to study the first-epoch of AGN activity, including the formation of first massive blackholes; and (3) detect, measure, and map out molecular hydrogen rotational line emission from primordial cooling halos that are the formation sites of first stars and galaxies at $z > 10$. Related to (3), molecular hydrogen is now understood to be the main coolant of primordial gas leading to the formation of first galaxies. It is also the most abundant molecule in the universe. Molecular hydrogen cooling in primordial dark matter halos will likely be the primary tracer to study the transition from dark ages at z > 20, when no luminous sources exist, to reionization at z < 10. At a metallicity of $10^{-3.5}$ $Z_{solar}$ gas cooling will transit from $H_2$ to atomic fine-structure lines. At z < 8, when primordial molecular hydrogen is easily destroyed by UV radiation, the prevalence of shocks in the ISM may provide ways to form a second and later generations of molecular hydrogen. The rotational lines of molecular hydrogen span across a decade of rest wavelengths from 2 to 30 μm. The detection of molecular hydrogen cooling requires an exteremely sensitive telescope (line sensitivity down to $10^{-23}$ W m$^{-2}$) and will likely be the primary driver of spectral line sensitivity of the Far-IR Surveyor.

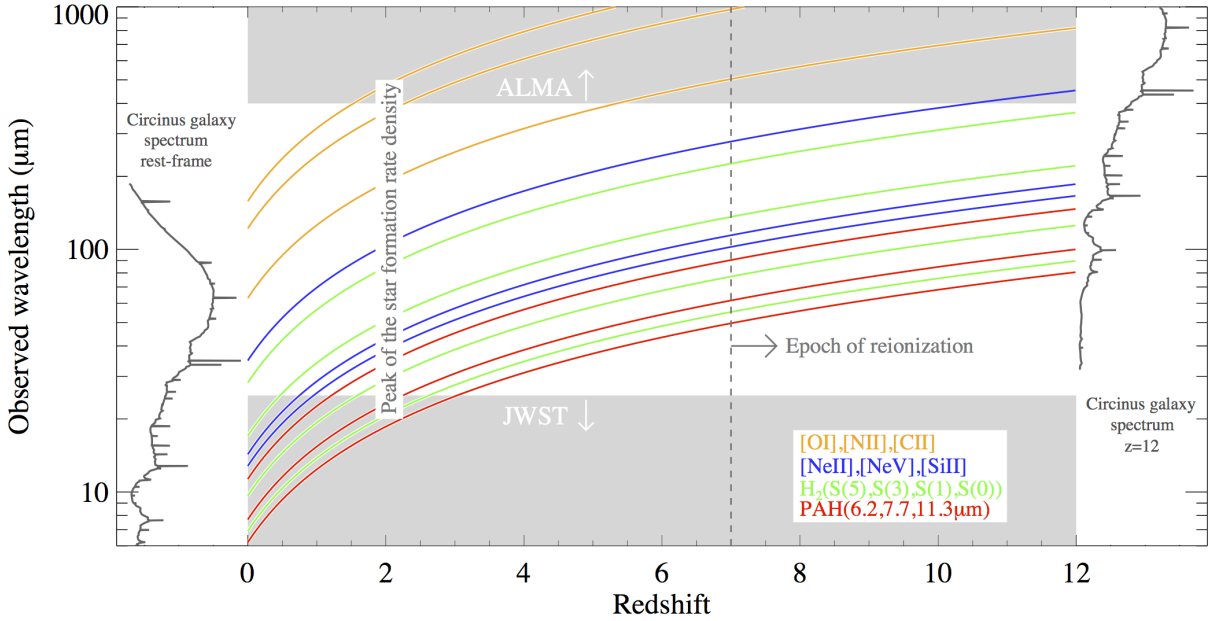

**Figure 1: Key MIR and FIR spectral lines for diagnosing star formation and black hole growth are uniquely shifted to the Far-IR Surveyor spectral coverage window over much of cosmic time.**

### 2.2 Evolution of Galaxies and Blackholes

With a huge leap in sensitivity and access to a rich astrophysical window of diagnostic features (Figure 1), the Far-IR Surveyor will uniquely address fundamental questions about how galaxies and black holes are born, how they evolve over cosmic time, and how gas, dust and metals are created and enrich the interstellar and intergalactic medium. With a large, cold telescope in space, the Far-IR Surveyor will be able to measure a wide variety of emission and absorption lines from atomic and molecular gas, and dust in thousands of galaxies, independent of the level of obscuration, providing our first, clear and unbiased picture of galaxy evolution from the local Universe to the epoch of Reionization. Through sensitive spectroscopy over the large range in wavelengths inaccessible to JWST and ALMA, the Far-IR Surveyor will be able to measure star formation rates and black hole accretion rates simultaneously. Spectral lines over this wavelength range also probe the build up of metals and the dynamic physical processes that regulate feedback and how galaxies grow over time. At early cosmic time, the Far-IR Surveyor can uniquely measure molecular hydrogen to trace turbulent formation of the first structures in the Universe. Through wide-area surveys, the Far-IR Surveyor will probe these processes in all physical environments, from the field to dense clusters, and provide a complete census of the dusty, hidden Universe.

### 2.3 Local Volume of Galaxies, Milky Way, Interstellar medium, Evolved stars and Star Formation

Studying the interstellar medium in the Milky Way and nearby galaxies provides the link between the formation of stars and planetary systems and the evolution of galaxies over cosmic time. Most of the fundamental diagnostic tools for such studies are found in the far-IR: the peak and long-wavelength tail of the dust spectral energy distribution (SED) and the dominant cooling lines for most ISM phases ([CII] 158 um, [OI] 63 um, [OIII] 88 um, [NII] 122 & 205 um, etc.). In addition to pushing into new regimes of sensitivity and angular resolution with these fundamental tracers of ISM energetics and physical state, the Far-IR Surveyor will provide an array of powerful new tools for studying the ISM and star formation. Lines such as HD and low-lying $H_2O$ rotational lines that were detected in only the brightest peaks of nearby star-forming regions like Orion with previous IR telescopes will become routine tracers of the ISM in the era of the Far-IR Surveyor.

The Far-IR contains key diagnostics for studying star formation and feedback across a wide-range of environments. From detailed studies of Galactic-disk star-forming regions, to feedback-driven galactic outflows in nearby galaxies, the Far-IR Surveyor will allow us to constrain how star formation and feedback are regulated in galaxies and the dependence of these processes on environment. Measurements of the peak of the SED for dust and far-IR cooling lines will enable radical progress in our understanding of star formation, in particular: (1) by measuring the amplitude and variability of accretion onto protostars with the far-IR SED to determine how they gain mass and what sets the initial mass function and (2) by measuring the phase structure of the ISM and tracing the cycling of gas between phases. The Far-IR uniquely enables detailed kinematic studies of the main cooling lines of the ISM and higher-energy molecular transitions, tracing the energetics of gas from protostellar cores through galactic winds. These measurements are critical for studying feedback processes in the local universe and their interactions with the wider environment.

The Far-IR is also host to a variety of diagnostics that trace the abundance and characteristics of dust and molecules in the interstellar medium. The evolution of dust in the ISM and the properties of molecular gas remain critical questions for studying galaxy evolution and star formation at all redshifts. The peak of the dust SED and the Rayleigh-Jeans tail at long wavelengths are critical for measuring dust masses. The Far-IR Surveyor will enable measurements of the dust SED and mass in the local ISM and nearby galaxies at very low surface brightness, probing environments where the dust life-cycle is very different from the local area of the Milky Way (i.e. low metallicity dwarf galaxies, the diffuse atomic outskirts of spiral galaxies). The far-IR will also provide spectroscopic measurements of the long-wavelength dust emission to constrain the growth of grains in the Milky Way molecular clouds. The greatly enhanced sensitivity of the Far-IR Surveyor compared to other infrared telescopes also provides access to unique tracers of the molecular ISM, which have never been systematically studied. These include the far-IR rotational lines of the HD molecule, which provides a unique handle on tracing molecular gas masses in the Milky Way and nearby galaxies; the high-J rotational lines of CO, which trace energetics in shocked or highly irradiated molecular gas; and the low-lying $H_2O$ lines, which can track the formation of $H_2O$ in the diffuse ISM through its incorporation into proto-planetary disks and eventually planets.

### 2.4 Protoplanetary Disks and Exoplanets

A next-generation mid- to far-infrared space observatory will uniquely answer questions fundamental to our understanding of planet formation and of how ingredients for life are delivered and incorporated into habitable planets. We aim to measure the total gas mass across all evolutionary stages of planet formation in disks across the entire stellar mass range, using the strong ground-state line of deuterated molecular hydrogen, HD at 112 um. Only a single unambiguous gas-mass measurement of a protoplanetary disk was made with Herschel, while a cold far-infrared space telescope will be able to measure the mass of thousands of disks. The Far-IR Surveyor will create a comprehensive census and mass inventory of both water vapor and water ice, as well as other volatile molecules, in planet-forming regions, using the 43 um water ice band, and a multitude of far-infrared water lines. The power of a Far-IR Surveyor can be leveraged to accurately measure the total water content, and its spatial distribution, in disks around stars of all masses (including brown dwarfs) and at distances large enough to include massive star-forming regions, similar to those our own Sun formed in.

The Far-IR Surveyor will also address fundamental questions on how planetary systems evolve after they initially form. It will determine the true frequency of Kuiper-belt analogs and other planetesimal systems around all stellar masses. It will also measure the mineral and volatile composition of typical debris disks. It will determine the frequency of wide-orbit ice giants (Uranus and Neptune analogs) by imaging sculpted debris disk structure.

The Far-IR Surveyor will be able to detect the thermal emission peak of cool exoplanets at habitable temperatures (~300 K) at wavelengths where the star-planet contrast is most favorable, and measure their atmospheric composition. This may be accomplished by a combination of transit spectroscopy, eclipse spectroscopy, and direct coronagraphic imaging. Important atmospheric diagnostics include spectral bands due to ammonia from 10-50 um (a unique tracer of nitrogen), the 15 um $CO_2$ band (an important greenhouse gas), and many water bands.

### 2.5 Solar System

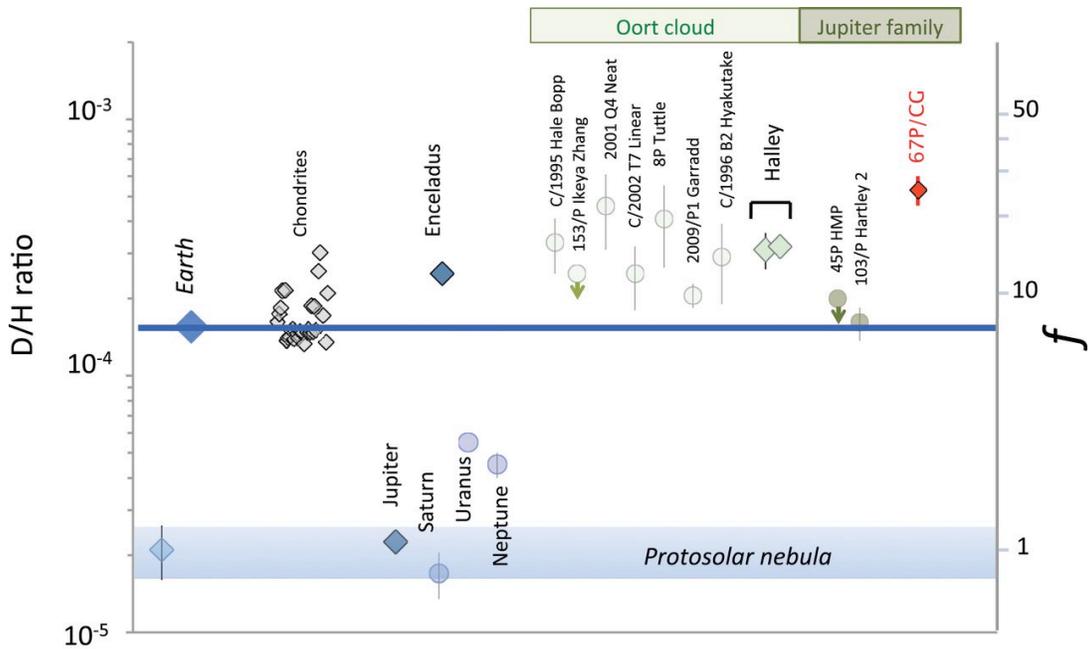

**Figure 2: The D/H ratio measured in comets compared to the planets and protosolar nebula, from Altwegg[10].The Far-IR surveyor would add 1000s more data points to enable a comprehensive study of the conditions of our early solar system.**

A realization of how processes that form planets and create habitable conditions is through studies of our own solar system. However, the events that sculpted its particular morphology are not well constrained, and current theories, such as the Nice and Grand Tack models, still await detailed measurements of the outer solar system. Isotopic ratios serve as cosmogonic "thermometers" that trace physical variation across the disk and small body populations in the young solar system, and may provide insight to interstellar inheritance and the degree of processing that may have occurred.

Recent measurements of the D/H ratio in comets have varied from solar to enriched values in deuterium (Figure 2). The limited sample needs to be increased, and can be probed in the far-IR through measurements of water in a much larger sample of comets (periodic and long period) with a new space-observatory that is significantly more sensitive to previous facilities. Measurements of other isotope ratios in carbon, oxygen, and nitrogen can accurately determine the primordial conditions of our Solar System, and probe the origins of the small body populations.

Additionally, the far-IR provides access to other tracers for the history and evolution of the solar system. The thermal history can be sampled through detailed studies of Hydrogen and Helium in the giant planets (Jupiter, Saturn, Uranus, Neptune). Thermal measurements of the outer solar system small bodies (TNOs, Kuiper Belt objects, etc.) can only be conducted in the far-IR (where we can measure 1000s of objects) and will provide better constraints on the size distribution of these targets and insight into their composition and evolution.

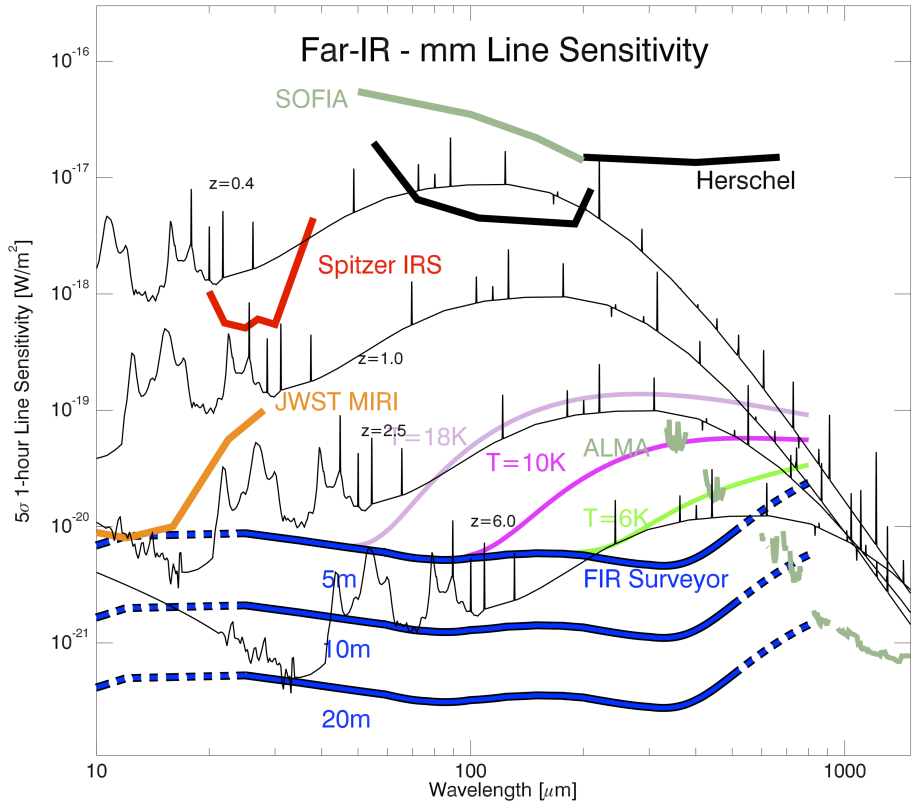

**Figure 3: The potential Far-IR-mm line sensitivity of a Far-IR Surveyor with respect to recent, current and near-future observatories. The template spectral energy distributions for an active star forming galaxy at different redshifts provide a scientific context for the gain in sensitivity. The blue lines show the sensitivity for a telescope system cooled to 4 K with collecting areas equivalent to a single aperture with a diameter of 5, 10 or 20 m. For the 5 m case, we also show the effects of increasing the temperature of the mirrors to 6, 10 or 18 K that motivates the need for cold optics.**

## 3.  NEW SCIENCE NEEDS NEW TECHNOLOGY

Several technologies will need development in order to make a Far-IR Surveyor a reality (Table). Aperture options include a large filled aperture or an interferometer.  For either, the most important technology is large arrays of direct detectors which are sufficiently sensitive to meet the zodiacal-light photon background, per-pixel sensitivities need to be on order $3 \times 10^{-20}$ W Hz$^{-1/2}$ for spectroscopy.  These sensitivities have been obtained with transition-edge-sensed (TES) bolometers, as well as quantum capacitance detectors (QCDs).   However, work is required to create a detector system should be compatible with formats of up to 1 million pixels in the full observatory, this requires high-density (~1000 x) frequency-domain multiplexing as is now employed with the kinetic inductance detectors (KIDs).    Everything needs to be cold with detectors at ~50 mK and the telescope optics ~4 K, both can be obtained by scaling up existing cryocooler systems, sub-Kelvin coolers, and space observatory thermal design.  Instrument technologies such as compact direct-detection spectrometers are also needed to enable the large format spectroscopy envisioned for either the single-aperture or interferometer concepts.

While it does not take advantage of the cold telescope, heterodyne spectroscopy is also a key scientific tool, and arrays of ~100 heterodyne receiver pixels with sensitivity approaching the quantum limit are also important.   Finally, we note that cryogenic actuators are needed and materials properties (e.g., damping, emissivity, structures) are important, and must be measured.

| New Technology | New Capability |
| --- | --- |
| Space | Wavelength coverage JWST<—>ALMA |
| Cold Mirror | Continuum & Spectroscopic line sensitivity |
| Large Telescope | Spatial resolution and sensitivity |
| Large Detector Arrays | Wide field imaging |
| Compact Gratings & Integrated Spectrometers | 3D mapping |
| Mid-IR Coronagraph | Exoplanet+Disk Characterization |

## 4.  FUTURE PLANS

The Far-Infrared Surveyor Mission Concept study for the 2020 Decadal Survey is a community driven exercise. Such a large mission will need international participation and support to make it a reality. Moreover, partnerships with industry are essential to the study, proposal and building a Far-Infrared Surveyor. The STDT will be consulting with international partners, industry in addition to the astronomical community and the NASA labs during the study.

This summer/fall 2016, the STDT is working with the larger astronomical community to create key science question proposals. The STDT will review all the questions to identify the driving questions that will define the science requirements for the observatory. They plan to identify an aperture architecture (e.g. filled or interferometer) based on science before commencing on the serious engineering study for the mission concept. In 2017, the STDT will work with the engineering team to develop the mission concept with engineering studies and independent costing. In 2018, the STDT will finalize the concept and science case. In March 2019, the community co-chairs will present the report to the Decadal committee.